# Bipartite Synthesis Method applied to the Subset Sum Problem demonstrates capability as decision and optimization tool

## Abstract

This paper introduces a deterministic algorithm for solving an instance of the Subset Sum Problem based on a new method entitled the Bipartite Synthesis Method. The algorithm is described and shown to have worst-case limiting performance over similar to the best deterministic algorithms achieving run time complexity on the order of $O(2^{0.5n})$. This algorithm is representative of a more expansive capability that might convey significant advantages over existing deterministic or probabilistic methods, and it is amenable to blending with existing methods. The method introduced can be applied to a variety of decision and optimization problems.

## Introduction

The Subset Sum Problem (SSP) is one of the recognized Non-Deterministic Polynomial time Complete (NPC) problems of importance to the field of computational optimization and data analysis [1]. Typical of NPC problems, there is a critical value of a parameter that delineates a region where problems are very easy to solve. This transitions rapidly to a hard region where no algorithms perform efficiently [2]. All problems within this class are defined by the presence of at least one problem instance that generates exponential run time growth with input size for any known algorithm.

In the case of the SSP, a naive search requires on the order of $n2^n$ combinations to find all solutions. This brute force approach has been improved upon by a process that splits and sorts the set of numbers achieving $O(2^{n/2})$ [3]. This is broadly recognized as the best performance in terms of deterministic algorithms. Deterministic algorithms have value in that they perform consistently, and improvements that provide improved efficiency for majorities of problem instances can complement increasing hardware processing speeds to accomplish more tasks.

Probabilistic algorithms are in many cases more efficient at solving decision problems, but they produce results with a small probability that the program will fail. For example, a statistical algorithm to determine if a particular SSP solution is likely to exist produces results with steps better than proportional to $2^{n/3}$ for worst-case scenarios [4]. The probabilistic search approaches rely on pseudo-random number generators, which do not always perform as intended. Rarely, one might not find the correct answer or only after a very long time. The uncertainty associated with their performance could be a basis for discounting their value.

A new deterministic algorithm has been devised for solving the SSP at least as efficiently as the best deterministic algorithm. Because it was developed based on a novel graph theoretical method it has features that might convey advantages to it over other algorithms. This paper is solely intended to describe the algorithm and show that it has good run time efficiency in the worst case compared with other deterministic algorithms. Although some testing of the algorithm has been made with random inputs, describing the performance exhaustively and comparing it to other algorithms is beyond the scope.

The particular SSP chosen for this paper requires the enumeration of any and all subsets $S'$ of a positive integer set $S$ that sum to a target value $t$. A simple example is finding subsets of the integer set {2, 5, 6, 7, 8} that sum to nine. Because this set is small, the single subset solution is easily found: {2, 7}. All solutions to this SSP can be placed in the following notation:
$$\Sigma\, \epsilon_i\, a_i = t$$

where $a_i$ is an integer contained within the set of $S$ integers, which can also include duplicate numbers. The program developed from the algorithm can apply both kinds of sets, unique integers and potential for repeat values within the set. For the particular SSP to be described, the factor $\epsilon_i$ is either zero or one. The equation is a constrained problem representative of the more general class of knapsack problems.

# Algorithm Description

The approach applies a technique of bipartite synthesis, a patent-pending method introduced in this paper.[1] The Bipartite Synthesis Matrix (BSM) is an extension of the bipartite graph applied to integer data to produce a unifying data structure. A bipartite graph is represented mathematically as a rectangular matrix relating two kinds of data, a set of elements arrayed against attributes. The BSM incorporates intervals in place of attributes. The intervals interrelate data elements through shared attributes and quantify the extent of relationships.

The process of applying the BSM is reminiscent of the bucket sort technique. The bucket sort technique applies non-overlapping intervals spanning a range of values as the basis for establishing a sorting process. The unsorted set is read in and each number assigned to an appropriate bucket. These buckets are of increasing value. The bucket intervals are organized in a way that generates an ordered process when the intervals are read out sequentially. The size of the buckets can be scaled to smaller and smaller sizes to increase the granularity of the sort.

There are several distinguishing features of the BSM method as opposed to the bucket sort. The intervals of the BSM can be used to sort, but in this instance they are re-purposed to evaluate solutions to the SSP. The process determines if combinations of intervals meet certain rules. The intervals are associated with coefficients. The coefficients and their respective intervals are constrained by the requirement that the combination of intervals and coefficients meet certain conditions defined by a target value to include or exclude various possible solutions.

The intervals enable multiple combinations to be evaluated simultaneously. If a few, large intervals are applied initially, many combinations of data elements can be checked using coefficients that equate to subset size. Relatively few combinations of coefficients are required for a few intervals. The successful combinations of coefficients and intervals transition to more coefficients and finer intervals. The goal is to filter more combinations faster through constraints than the coefficient combinations are allowed to grow.

The most fundamental example of a BSM is one in which an array S of n data elements is related by a single, closed interval [1,R] spanning some integer range 1 to R such that $S \in [1, R]$. Each data element is assigned to the interval. A coefficient C is also associated with the interval. The coefficient value can be any integer 1 to n equaling the size of a specified subset S'. For the initial interval, the number of potential coefficients to be evaluated includes all subset sizes. This generates some number of coefficients L equal to the set size.

The process of applying constraints to the single interval and its range of coefficients is not completely trivial. In this first stage the population of data elements contained within the interval must satisfy the constraint that it be greater than or equal to the coefficient. Although that is trivial, what is not trivial is testing the boundaries of the interval to determine if solutions can exist within. This process applies interval arithmetic to determine if the interval envelopes some target value t:

$$C * [1,R] \ni t$$

The above statement is evaluated by multiplying a coefficient C with the lower interval limit to determine if the resultant interval is less than or equal to the target value. Similarly, C is multiplied with the upper limit to verify that the product equals or exceeds the target value. If true, then a potential solution can exist. The coefficients that satisfy the constraints are retained for the next stage and those that fail one or both constraints are discarded.

A multi-scale approach is used to expand the process from one to multiple, smaller intervals. The process doubles the number of intervals within the range for successive stages until the unit scale is achieved. Two intervals, [1, a] and [a+1,R], are created from the original interval. Because R is an integer, it is convenient to make it some value of $2^x$ greater than the maximum of S to facilitate the integer division. Each coefficient from the previous stage is used to develop one or more combinations of coefficient pairs for this stage.

For each interval combination every interval is assigned a coefficient creating various coefficient pairs. These new coefficients are children of a single coefficient that pass initial constraint tests. The first constraint is generated by the requirement that each coefficient pair $C_1$ and $C_2$ must sum to the parent coefficient. Multiple combinations can be made stemming from meeting this equation:

---

[1] Also see USPTO Application Number 14052288 entitled Multivariate Data Analysis Method.

$$C = C_1 + C_2$$

It can be found that the maximum number of child coefficient combinations for any value of C passed down according to the above constraint is equal to C.

The population test is applied to each of the coefficient combinations, where it is possible to have population distributions in child intervals inconsistent with the values of their respective coefficients.

The second constraint is related to the summation target. The target constraint test can also be applied to the interval pair using the coefficient pair:

$$C_1 * [1,a] + C_2 *[a+1, R] \ni t$$

The coefficient combinations that fail either constraint test are discarded.

The process of repeatedly doubling the number of intervals within a range by halving their size can continue until no coefficient combinations remain or if the unit interval scale (degeneracy) is achieved, after which the solution or solutions are read out. The presence of solutions at the unit scale answers the decision problem.

This problem naturally lends itself to one of optimization. If no coefficient combinations are reached at the unit scale, it is possible to evaluate for optimal solutions by using coarser scale solutions. From the last stage that contains coefficients that meet the constraints, solutions can be taken and examined individually to determine if one or more solutions represent the optimal solution, one closest to the value of the target.

## Algorithm Implementation and Proof of Correctness

A rendition of the program in pseudocode is as follows:

```
for i = 1 to Set_Size
    j = Init_Scale                          // some initial value 2^x ≥ max set integer value
    Nr_Intervals = 1                        //initialize partition number
    while j ≥ 1 do
        for k = 1 to Nr_Intervals
            Lower_ Interval_Limit(k)        //Procedure returns LL_Interval[k]
            Upper_ Interval_Limit(k)        //Procedure returns UL_Interval[k]
            Calc_Pop_BSM( k)                //Stores BSM_Population[k]  (ie. sum of all instances)
        end 'for k'
        // Make Input Matrix of seed coefficients to produce new coefficient table
        Generate_Coeff_Input_Table          // Coeff_Input_Table made from  previous
                                            // Interim_Solutions_Table (or initialized)
        for each Coeff_Input_Table row do
            Build_Coeff_Table               // Procedure makes Coeff_Table from  Coeff_Input_Table
            for l = 1 to Coeff_Table_Size   //# of rows determined in Build_Coeff_Table proc.
                for k = 1 to Nr_Intervals
                    if BSM_Population[k] < Coefficient_Table[k,l]
                        Test = 'false'
                        end 'for k'
                    end if
                    Sum_LL = Sum_LL + Coefficient_Table[k,l] * LL_ Interval(k)
                    Sum_UL = Sum_UL + Coefficient_Table[k,l] * UL_ Interval(k)
                end 'for k'
                if Sum_LL > Target then Test = 'false'
```

```
            if Sum_UL <  Target then Test = 'false'
             if Test ! = 'false' then Output_Coeff_Table_to_Interim_Solutions_Table
        end 'for l'
          if j == 1 then Solutions_Table   = array_merge (Solutions_Table, Interim_Solutions_Table)
     end 'for each Coeff_Input_Table row'
     j = j / 2                                              //more granular scale
     Nr_Intervals  =  Init_Scale / j                        //double the number of intervals mapped in range
  end 'while'
  Print Solutions_Table
end 'for i'
```

There are two primary loops, one iterating over the number of integers in the set and the second expanding the granularity from coarse to unit-scale size. It is possible to economize memory storage requirements by sequentially processing a single coefficient input into a table of coefficient combinations and testing with the constraints. True values that successfully pass all of the constraint tests are stored in a new solutions table. The Coefficient Combination table is cleared and the process repeated. This is not explicitly shown on the process diagram.

During the multi-scale process R is repeatedly halved. The number of intervals thus doubles with each cycle. Effectively, each interval pair, or branch of a binary tree, is created by dividing an interval through its midpoint into two child intervals. One child acquires the lower boundary and the midpoint value of the parent. The second child receives the midpoint value plus one as its lower boundary along with the upper boundary of the parent. This process can repeatedly take place until unit scale is achieved and there are R intervals. At this point the intervals are effectively degenerate being at most one integer in diameter and containing at most one value.

The process is ended if any of a number of conditions occur. The process can end after degenerate intervals are reached for every subset size. The process might end earlier if solutions are found (assuming that was the goal). If no solutions exist, the solutions table remains empty. If no combinations of coefficients exist with respect to a set of intervals at a coarser scale, then the process might also stop well before degeneracy is reached.

An important feature of high density data sets is the generation of blocks of sequential values that can be expressed as intervals in their own right. These blocks of sequences, which can be hidden within a disordered set S, are discoverable within the BSM for sets with non-redundant values. The integer diameter of an interval equaling the population of the bucket interval indicates a sequence block.

If a child interval is discovered to contain a complete block of values, this value multiplied by an appropriate coefficient can be subtracted from the summation target value to achieve a reduced target interval. This reduced target interval can be used subsequently for evaluating the newly truncated instance of the problem. The result is fewer coefficients and thus fewer combinations that require testing.

In addition to the block intervals, the process can be discontinued in instances when the coefficient for an interval equals the population of elements within it. If for instance the coefficient is two and there are two elements, there is only one mode possible. This represents a coefficient-filled interval and is a constant value. The combinatorial process would stop for that particular coefficient-filled interval but not necessarily for the whole program.

Alternatively, the process is amenable to statistical approaches by evaluating the subsets and conditions that are more probable to produce solutions. This method could also be blended with other deterministic methods.

## Analysis of Run Time Complexity

The maximum run time expansion of the algorithm is based on estimating the maximum expansion of the coefficient combinations for the worst case situation. The evaluation of hard SSPs, data where the density $\rho$ approaches unity can be difficult to solve with the above process alone. Presuming an arbitrary target value, a worst case condition would be maximizing the number of descendant generations. The number of coefficients increases by recursively doubling as the coefficient values decrease by a factor of two. With a naive approach, a scenario can be envisioned whereby the process could take time proportionate to $2^n$ steps to complete after evaluating every combination and reaching the final, degenerate interval scale.

To evaluate the extreme growth, one must evaluate the worst case conditions to determine how early one can halt the process to avoid reaching the degeneracy stage when numbers of intervals are maximized. Near the final stage, interval blocks will occur as described above and intervals will become filled. These events can be used to define worst-case halting before degeneracy is reached.

Under worst case conditions, the algorithm can be halted before the final stage. The next-to-last possible scale in the algorithm is an interval scale of two. At scale two, any cluster can hold one or two data elements. If the population is two, regardless of the coefficient size being one or two, the interval is full. In this case, the interval multiplied by its coefficient can be subtracted from the SSP target value to achieve a new target that is now an interval. This effectively ends the iterative operation involving this interval.

Likewise, if the cluster only has a single data element, then the coefficient is restricted to a value of one, and the interval can be made degenerate given the value of the single element. This element can be propagated down or subtracted from the target; either way, it does not expand the combinatorial process.

Thus, the second-to-last generation of combinations can be avoided, and it is this third prior scale that presents the earliest stopping potential for the worst case scenario. Any interval of scale four can have a maximum of three data elements without terminating because of fullness. With three data elements, the coefficient cannot be three or four without terminating the evaluation. For a single element, the process can also be halted because of degeneracy. For the case of a coefficient value two with three data elements, this presents a requirement for three evaluations. Thus, the worst case exists at unit scale four when there remain n/3 intervals each having two elements. This event would generate $3^{n/3}$ evaluations. A simple conversion indicates that the worst case run time is on the order of $2^{0.5n}$.

The memory size complexity appears to be determined by the size of worst-case arrays produced. There is no sorting process and thus no inherent reason for storing combinations after testing. This suggests that the space complexity is no greater than the run time complexity and likely some polynomial relation of n.

## Conclusion
The Bipartite Synthesis Method was applied to the Subset Sum Problem. It exemplifies a new deterministic approach to computation. Given naive rules, the process improves against the brute force method in the worst case achieving $O(2^n)$ versus $O(n2^n)$ run time complexity. By application of some additional rules, the interval-based approach limits run time expansion to as good as the best deterministic method achieving a worst case run time complexity of $O(2^{0.5n})$.

Although the method does not offer a fundamental breakthrough in absolute run time complexity, the example does demonstrate that the method is capable of performing decision tasks with good efficiency. The multi-scale feature involving intervals indicates the inherent capability exists to parallelize the algorithm.

The interval-based method can be adapted to a range of decision and optimization problems. Even foregoing the performance against other mature algorithms, the simplicity and efficiency for real-world scenarios should not be discounted.